\documentclass[runningheads]{llncs}
\usepackage{amsmath} 
\usepackage{booktabs} 
\usepackage{xcolor}
\usepackage{dsfont}
\usepackage{graphicx}
\usepackage[capitalise]{cleveref}
\usepackage{listings}
\usepackage{tabularx}

\definecolor{ForestGreen}{rgb}{0.13, 0.55, 0.13}

\usepackage{url}

\begin{document}
\title{Non-Parametric Class Completeness Estimators for Collaborative Knowledge Graphs --- \\ The Case of Wikidata}

\titlerunning{Non-Parametric Class Completeness Estimators}
%
\author{Michael Luggen\inst{1}
 \and
 Djellel Difallah\inst{2}
 \and
 Cristina Sarasua\inst{3}
 \and
 Gianluca Demartini\inst{4}
 \and
 Philippe Cudr\'e-Mauroux\inst{1}
 }
\authorrunning{Luggen et al.}

\institute{University of Fribourg, Fribourg, Switzerland
 \email{\{firstname.lastname\}@unifr.ch}
 \and
 New York University, New York, USA
 \email{djellel@nyu.edu}
 \and
 University of Zurich, Zurich, Switzerland
 \email{sarasua@ifi.uzh.ch}
 \and
 University of Queensland, Brisbane, Australia
 \email{demartini@acm.org} 
 }

\maketitle              

\begin{abstract}
Collaborative Knowledge Graph platforms allow humans and automated scripts to collaborate in creating, updating and interlinking entities and facts. To ensure both the completeness of the data as well as a uniform coverage of the different topics, it is crucial to identify underrepresented classes in the Knowledge Graph.
In this paper, we tackle this problem by developing statistical techniques for class cardinality estimation in collaborative Knowledge Graph platforms. Our method is able to estimate the completeness of a class---as defined by a schema or ontology---hence can be used to answer questions such as ``Does the knowledge base have a complete list of all \{Beer Brands|Volcanos|Video Game Consoles\}?'' 
As a use-case, we focus on Wikidata, which poses unique challenges in terms of the size of its ontology, the number of users actively populating its graph, and its extremely dynamic nature. Our techniques are derived from species estimation and data-management methodologies, and are applied to the case of graphs and collaborative editing. In our empirical evaluation, we observe that i) the number and frequency of unique class instances drastically influence the performance of an estimator, ii) bursts of inserts cause some estimators to overestimate the true size of the class if they are not properly handled, and iii) one can effectively measure the convergence of a class towards its true size by considering the stability of an estimator against the number of available instances.
\end{abstract}

\section{Introduction}

Knowledge Graphs (KGs) play a critical role in several tasks including speech recognition, entity linking, relation extraction, semantic search, or fact-checking.
Wikidata \cite{VrandecicKroetzsch14cacm} is a free KG that is collaboratively curated and maintained by a large community of thousands of volunteers. With currently more than 55M data items and over 5.4K distinct properties that help describe these data items, Wikidata is the bridge between many  Wikimedia projects (e.g., Wikipedia, Wikimedia Commons, and Wiktionary), as well as the interlinking hub of many other Linked Data sources. Its data is consumed by end-user applications such as Google Search, Siri, and applications to browse scholarly information\footnote{Scholia \url{https://tools.wmflabs.org/scholia/}}.

Being a collaborative, crowdsourced effort, Wikidata's data is highly dynamic. Editors can create items individually (e.\,g.\, a new instance representing a natural disaster that just happened), or in bulk (e.\,g.\ importing data about all the pieces of art in a city) about any topic that satisfies the notability criteria defined by the community\footnote{Wikidata's Notability \url{https://www.wikidata.org/wiki/Wikidata:Notability}}. The open curation process leads to a KG evolving dynamically and at various speeds. While such a process is beneficial for data diversity and freshness, it does not guarantee the total (or even partial) \textit{completeness} of the data. Given that previous research has shown that data consumers identify completeness as one of the key data quality dimensions \cite{wangdataqual96}, together with accuracy and freshness, it is of utmost importance to provide mechanisms to measure and foster data completeness in collaborative KGs.

In that context, the Wikidata community has already endorsed a series of initiatives and tools that encourage efforts towards population completeness \cite{Zaveri2015:LODQ}. For instance, there are WikiProjects\footnote{Wikidata WikiProjects \url{https://www.wikidata.org/wiki/Wikidata:WikiProjects}} that aim at populating Wikidata with bibliographic references, genes, or notable women.

With such a decentralized approach of independently-run data entry and import efforts, it has become very difficult to understand and measure what is still missing in Wikidata. While there is related work that measures the completeness of item descriptions in Wikidata (see Section \ref{sec:relatedwork}), there is (to the best of our knowledge) no systematic approach to measure \textit{class completeness} other than by manually checking for candidate entities and facts to be inserted in the KG.

In this paper, we focus on the specific problem of \textit{estimating class completeness} in a collaborative KG and experimentally evaluate our methods over Wikidata.
We limit our work to the family of finite classes, where the number of instances in such classes is fixed.
We take a data-driven approach to that problem by leveraging models from statistics and ecology used to estimate the size of species \cite{chao1992estimating}. We propose methods to calculate the cardinality of classes and build estimates for the class convergence to the true value. We note that while we focus our empirical study on Wikidata, our proposed methodology is applicable to any other collaborative graph dataset with analogous characteristics, where the action log describing its evolution is available.
By calculating the expected class cardinality, we are able to measure class completeness given the number of instances currently present in the KG for that class.
We evaluate different class size estimation methods against classes whose sizes are known through trustworthy third-party sources (e.g., the number of municipalities in the Czech Republic) and for which a complete ground truth exists. We then apply these methods to other classes in order to generate completeness estimates for other parts of the KG.

The main contributions of this paper are as follows:
\begin{itemize}
    \item We show how the edit history of a KG can be used to inform statistical methods adapted from species estimators (Section \ref{sec:model});
    \item We evaluate the effectiveness of statistical methods to estimate the class size and KG completeness based on repeated sampling (Section \ref{sec:experiments});
    \item We provide tools to make Wikidata end-users (both human and applications) aware of the incompleteness of many subparts in Wikidata (Section \ref{sec:tools}).
\end{itemize}

\section{Related Work}
\label{sec:relatedwork}

\paragraph{Data Completeness In Knowledge Graphs}
is one of the most important data quality dimensions for Linked Data \cite{Zaveri2015:LODQ}; it has also been acknowledged as a key data quality indicator by the Wikidata community\footnote{Wikidata Quality RFC \url{https://www.wikidata.org/wiki/Wikidata:Requests_for_comment/Data_quality_framework_for_Wikidata}}.
Different data cleaning methods proposed by the research community have focused on different types of completeness. For example, ReCoin \cite{Balaraman:2018:Recoin} measures the relative completeness that item descriptions have, compared to other items of the same type. It keeps track of used properties  and encourages editors to add new statements and foster more homogeneous item descriptions. 
Gal\'arraga et al. \cite{Galarraga2017completeness} investigate different signals to predict the completeness of \emph{relations} in KGs.
The work of Soulet et al. \cite{soulet_representativeness_2018} introduces a method to estimate the lower bound of completeness in a KG. The completeness is estimated through the missing facts to reach a distribution according to Benfords Law. Kaffee et al. \cite{kaffeeS18wikidata} study label completeness across languages. The work by Wulczyn et al. \cite{Wulczyn:2016:GWA:2872427.2883077} encourages Wikipedia editors to write different language versions of existing articles. Tanon et al. \cite{DBLP:conf/semweb/TanonSRMW17} uses association rules to identify missing statements, while Darari et al. \cite{Darari:2018:CMR:3240924.3196248} provide means to describe and reason over RDF statement completeness. 
To complement these methods, in this paper we consider the problem of class completeness in the KG.

\paragraph{Cardinality Estimation In Databases}
Estimating the cardinality of a table in relational databases is key to query performance optimization. This requires a combination of database technology and statistical methods and allows to compute the cost of database operations that are then used for optimization strategies (e.g., storage allocation and data distribution) \cite{Mannino:1988:SPE:62061.62063}.
Similarly, cardinality estimation is key to optimize query execution in RDF triplestores. The key difference with relational databases is the presence of many self-joins in queries over RDF data. This requires custom cardinality estimation techniques for SPARQL queries over RDF data \cite{neumann2011characteristic}.
In distributed databases, cardinality estimation is also a necessary step to optimize query execution. The key aspect is estimating the size of non-materialized views in a way that is accurate and provides statistical bounds \cite{papapetrou2010cardinality}.
Our work addresses the different problem of determining the cardinality of a class in a KG leveraging its edit history.

\paragraph{Data Completeness in Crowdsourcing}
The problem of counting items and individuals also arises in a crowdsourcing setting. 
Previous work \cite{difallah2018demographics} developed models to estimate the size of the crowd in Amazon MTurk by taking into account the propensity of a worker to participate in an online survey or micro-tasks, respectively. That work used \textit{capture-recapture}, a technique based on repeated observations of the same worker participating in tasks. In our class size estimation method, we estimate the size of data (not crowds) based on observations made through another form of crowdsourcing, i.e., volunteering.

In a similar setting, Trushkowsky et al. \cite{trushkowsky2013crowdsourced} tackled the problem of enumerating the list of all instances in a specific class through paid crowdsourcing. The crowd workers were explicitly asked to provide a list of distinct items, for example, ``input the list of all ice cream flavors''. Similar to our work, the authors used \textit{capture-recapture} techniques but also had to deal with aspects unique to a crowdsourcing environment. For instance, they introduced a ``pay-as-you-go'' method to estimate the cost-benefit ratio of crowdsourcing additional tasks to complement the current list.
They looked at both open-world and closed-world assumptions where the cardinality of the set is either known (e.g., ``list of US states'') or unknown and possibly unbounded (e.g, ``ice cream flavors'').
Their methods are based on techniques borrowed from ecology research to count the number of animals of a certain species, which we describe next.

\paragraph{Species Richness Methods}
In the field of ecology and bio-statistics, several capture-recapture techniques have been proposed to estimate the number of existing species \cite{bunge1993estimating,walther1998comparative}.
The idea of capture-recapture is to draw a sample at random from a population and to estimate the number of unobserved items based on the frequency of the observed items. Such approaches work well for closed populations, but different techniques are required when we allow for open populations. Open vs. closed population problems have fundamentally different questions to answer. The former focus on estimating the rates of arrival and departure, the latter is about size and propensity of capture. We restrict our work to the realm of closed classes since it was shown that if a closed population method is utilized when in fact there is a process of arrival/departure, then closed estimators tend to overestimate.
For example, the open-world-safe estimator ``Chao92'' \cite{chao1992estimating}  provides more accurate estimations when more evidence is available from a  class. We present our  results based on this and other estimators in Section \ref{sec:experiments}.

In our work, we look at the problem of estimating the size of a given class (e.g., Volcanos) or composite classes (e.g., Paintings drawn by Vincent van Gogh) in Wikidata. We tap into the edit patterns of Wikidata volunteers \cite{sarasua2018}, and apply capture-recapture techniques to estimate the completeness of a given class.

\vfill\clearpage

\section{Class Completeness Estimators}
\label{sec:model}

In this section, we introduce the family of estimators we leverage to tackle the class estimation problem in collaborative KGs.
First, we introduce the problem statement and the assumptions that we make in the context of Wikidata by defining the notion of \emph{class} in Wikidata.
Next, we introduce several statistical estimators ordered by complexity and show how they build upon each other.
In this paper, we refer to entities as all the instances of a particular class e.g., ``Cathedrals in Mexico''.

\subsection{Problem Definition}
\label{sec:problem}
Given a finite class $C$ of instances $I_{C}=\{I_1, ..., I_N\}$, our goal is to estimate the number of instances of $C$ i.e., $N=|I_{C}|$. We note $D$ the current count of instances of a given class in the knowledge graph. A class is complete once $D$ is equal to the true class size $N$.

The capture-recapture data collection protocol that we follow is based on $n$ observations recorded during $k$ successive \emph{sample periods}. 
Each observation relates to a direct or indirect change made to an instance of a specific class during the sample period (i.e., one month).
In practice, we extract \emph{mentions} from the edits in the knowledge graph. An edit is a change that either adds, modifies or deletes a statement involving one or more entities.
Every reference of an entity in the \emph{subject} or \emph{object} position of a statement defines a mention for the class that the mentioned entity belongs to.
In the end, each mention is composed of an entity (also called \emph{instance} because it belongs to a class), the class the instance belongs to, and a timestamp.

\subsection{Interpreting Edit Activity Related to Classes}
\label{sec:edits}
Given the edit history of a KG, we extract mentions as described in Listing \ref{lst:mention}: For every edit, we create a mention if one of the entities referenced belongs to a class. This is done on a per class basis.

\begin{lstlisting}[caption={Query on the Wikidata Graph illustrating the relation between edits and mentions on the example of the Single Domain class \emph{City} (Q515). (The properties referenced with the \emph{edit} prefix are not available in the public Wikidata endpoint.)},label={lst:mention},language=SPARQL]
SELECT ?instance ?timestamp
WHERE { ?instance wdt:P31/wdt:P279* wd:Q515.
        { ?mention edit:subject ?instance. }
        UNION
        { ?mention edit:object  ?instance. }
        ?mention edit:timestamp ?timestamp.          
}
\end{lstlisting}
 
\begin{figure}[tb]
\begin{center}
\includegraphics[width=.8\textwidth]{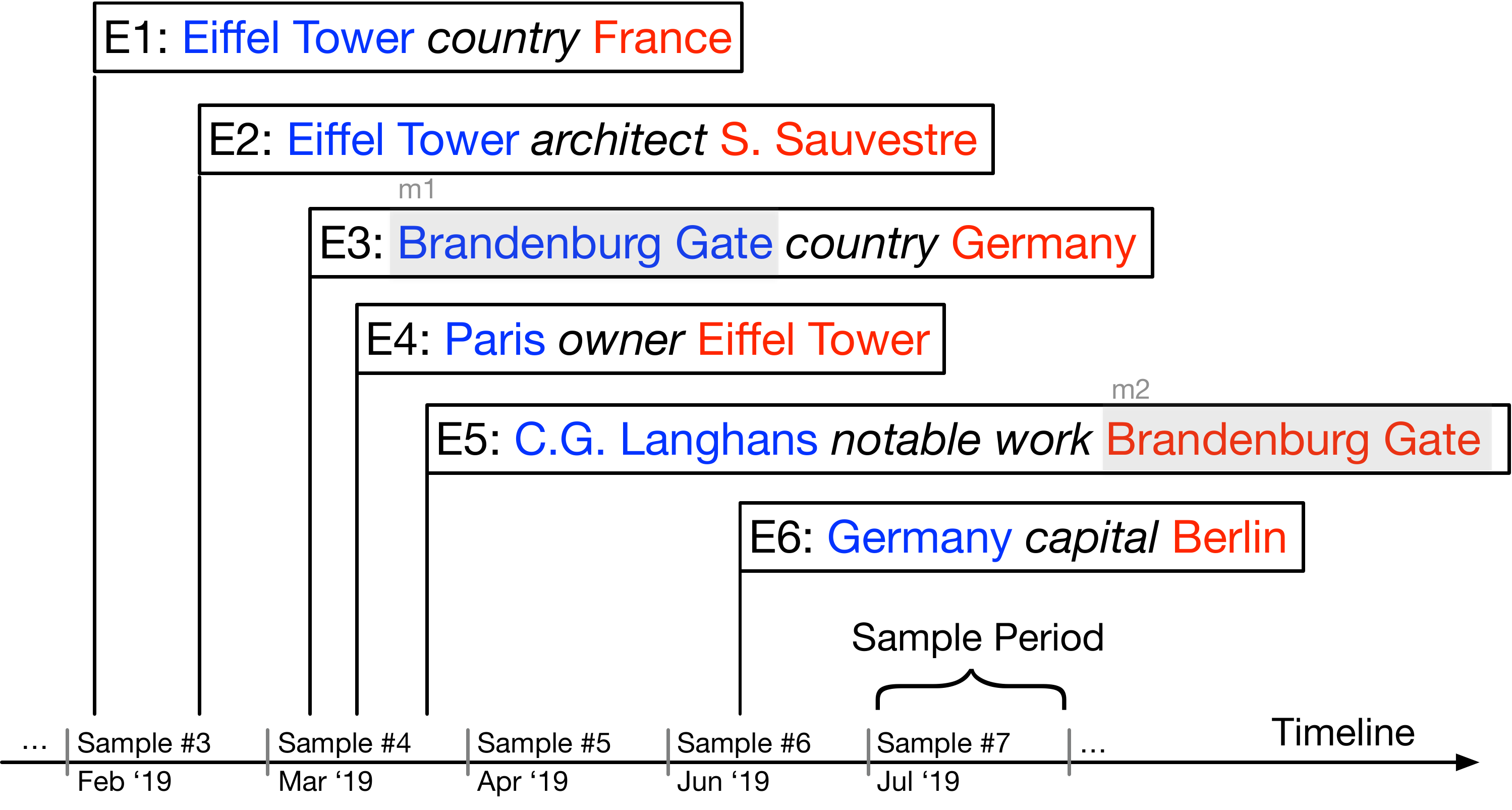}
\end{center}
\caption{The edits ($E_i$) of the Knowledge Graph (representing new edges) are leveraged to identify mentions. The source and target of each edge are collected to create a mention from the entity involved. Sample period \#4 contains 3 \emph{edits}, in which we identify 6 \emph{mentions}, from which we extract 2 \emph{observations} for class monument (despite the 3 mentions of entities of that class because $m_1$ and $m_2$ are only counted once), 1 observation for class country, 1 observation for class city and 1 observation for class person.}
\label{fig:timeline}
\end{figure}

We show in Figure \ref{fig:timeline} how the mentions get aggregated per sample period on the overall timeline.
In a given sample period, we count one observation per instance having at least one mention.
With $X_i$ being the frequency of observations relating to instance $I_i$, we compute the frequency of frequencies $f_i = \sum_{j=1}^{N} \mathds{1}[X_j = i]$, for $1 \leq i \leq k$. For example, $f_1$ is the number of instances observed only once (singletons), $f_2$ is the number of instances observed twice (doubletons) etc. With this notation, $f_0$ represents the number of instances that we never observed and we seek to estimate.
Each instance $I_i \in I_{C}$ of a given class has a unique probability $p_i$ of being mentioned, with $\sum p_i = 1$.

To be able to leverage the statistical techniques described below, the distribution of classes among the observations is supposed to follow a stationary multinomial distribution with unknown parameter $p_1, ... , p_N$. This leads to the following assumptions:
\begin{enumerate}
    \item The classes of interest are closed and countable as of the beginning of the experiment;
    \item The observations are independent events;
    \item In a class, the observations are at random and ``with replacement'';
    \item The probability of observing an instance within a class does not change over time.
\end{enumerate}

First, by assuming that classes are closed, we reduce the scope of the questions we can answer. Tracking the changes (growth and shrinkage) of an open class such as ``Events in Paris'' or ``Sitting Presidents'' would require a different approach, data, and assumptions.
Second, using a large number of edits made by different volunteers and scripts introduces a number of corner cases that we need to work with. It is for example possible to observe dependant actions, for example: systematically adding ``Name" followed by ``Date of Birth" or editors correcting each other. While our assumption is a simplifying one, we have not observed any significant correlations in the edits. This stems from the fact that the volunteers are not restricted on which edits they perform and what entities or classes they need to focus on.
The third assumption comes in contrast to the work in \cite{trushkowsky2013crowdsourced} where crowd workers were asked to list items that belong to a particular class.
Hence, a given crowd worker is answering from a population of possible items (i.e., sampling ``without replacement'').
In our case, Wikidata editors can create edits which repeatedly mention the same entity in the context of their work.
Finally, the fourth assumption is based on the fact that the observations we make are created through indirect references and are not directly related to the classes themselves.

\subsection{Non-parametric Estimators}
The intuition behind the estimators that we consider is based on the frequency of edits involving entities of a given class. To estimate the true class size we consider non-parametric methods that primarily use the frequencies of observations among instances.
Non-parametric methods do not assume any probability distribution of $p_i$ among the instances of a given class.

\subsubsection{Jackknife Estimators [Jack1]}

Jackknife (or ``leave-one-out'') methods have been used to develop species richness estimators \cite{heltshe1983estimating}. Similarly to k-fold cross validation, we use observations from $k-1$ periods by removing observations from one sample period from the data at a time and average the resulting pseudo-estimates made on each sub-fold.
We write $f_1^i$ to denote the instances observed only once in period $i$. We note that the number of distinct elements obtained when dropping period $i$ becomes $D_{-i} = D - f_1^i$.

We compute a pseudo estimate for each sub-sample of observations obtained by dropping the $i-th$ period using $\hat{N}_{-i} = kD - (k-1) D_{-i}$, and averaging across $k$.
A closed form of the first and second order Jackknife estimators is given by \cref{eq:j1} and \cref{eq:j2} respectively \cite{burnham1979robust}. 
We observe that $\hat{N}_{\textsc{jack1}}$ implies that the number of unseen instances is approximately the same as the number of singletons after a large number of sampling periods.

\begin{equation}
    \label{eq:j1}
    \hat{N}_{\textsc{jack1}} = D + \frac{k-1}{k} f_1
\end{equation}

\begin{equation}
    \label{eq:j2}
    \hat{N}_{\textsc{jack2}} = D + \frac{2k -3}{k} f_1 -
    \frac{(k-2)^2}{k (k-1)} f_2
\end{equation}

\subsubsection{Sample Coverage and the Good-Turing Estimator [N1-UNIF]}
The following methods are based on the concept of \emph{sample coverage} which is a measure of sample completeness.
\begin{equation}
    S = \sum_{}^{N} p_i \mathds{1}{[X_i > 0]}
\end{equation}
Since the probabilities of observing the instances as well as the population size are unknown, a popular estimate of the sample coverage is given by the Good-Turing Estimator~\cite{good1953population} \cref{eq:turing}. Effectively, this estimator relies on the complement of the ratio of singletons among the sample data and as an indicator of true sample coverage.
For example, if in past sample periods we have seen each instance only once, the probability of observing a new instance by collecting a new sample is $1$. Conversely, if all the instances were seen more than once, i.e., $f_1 = 0$ the probability of seeing a new instance in a new sample is reduced to $0$.

\begin{equation}
    \label{eq:turing}
    \hat{S} = 1 - \frac{f_1}{n} 
\end{equation}

If all instances have the same probability of being observed, the population size using the Good-Turing sample coverage is given by:
\begin{equation}
    \label{eq:uni}
    \hat{N}_{\textsc{n1-unif}} = \frac{D}{\hat{S}} =  \frac{D}{1 - \frac{f_1}{n} }
\end{equation}

We draw the attention of the reader to the trade-off that singletons and popular instances create. Typically, frequency counts will be heavily unbalanced and will tend to over or under-estimate the true population size. 

\subsubsection{Singleton Outliers Reduction [SOR]}
To mitigate the effect of the singletons on a class, a popular approach is to threshold the number of singleton elements.
Trushkowsky et al.~\cite{trushkowsky2013crowdsourced} proposed to limit the number of singletons introduced by a given contributor to two standard deviations above the mean of singletons introduces by other workers.
We adapt this method to our scenario by limiting the $f_1$ count to fall within two standard deviations above the mean.
The rationale behind our choice is to strike a balance between low and high dispersion of $f_1$ frequencies with respect to the set F of all frequencies that we observe.
\begin{equation}
    \hat{N}_{\textsc{SOR}} = \frac{D}{1 - \frac{\tilde{f}_1}{n}}
\end{equation}
with, 
\begin{equation}
\begin{aligned}
\tilde{f}_1 &= \min\Bigl\{f_1, 2\sigma + \mu\Bigr\} \\
\mu &= \sum\limits_{\forall j>1}^F \frac{f_j}{|F| - 1} \\
\sigma &= \sqrt{ \sum\limits_{\forall j>1}^F \frac{(f_j- \mu)^2}{|F| - 2} }
\end{aligned}
\end{equation}

\subsubsection{Abundance-based Coverage Estimator [Chao92]}
The work by Chao and Shen \cite{chao1992estimating} (hereon \emph{chao92}) uses the concept of sample coverage introduced above
and assumes that the probabilities of capture can be summarized by their mean i.e., $\bar{p} = \sum p_i / N = 1/N$ and their coefficient of variation (or $\gamma$) with $\gamma^2 = [N^{-1}\sum_i (p_i - \bar{p_i})^2]/\bar{p}^2$.

However, since we do not have access to the probabilities $p_i$ and $N$, the coefficient of variation is in turn estimated by using $\hat{N}_{\textsc{unif}}$ (via the Good-Turing estimator of sample coverage), and $p_i$'s with the observed data and corresponding $f_i$. 
\begin{equation}
    \gamma^2 =
    \max\Bigl\{
    \hat{N}_{\textsc{unif}} \sum_{i=1}^k \frac{i(i-1)f_i}{[n(n-1)]-1} , 0 
    \Bigr\}
\end{equation}

The \emph{chao92} estimator is given by \cref{eq:chao}. We note that if the coefficient of variation is close the zero, the estimator reduces to \cref{eq:uni} indicating an equiprobable scenario. Conversely, as $\gamma$ grows, signaling more variability in the probabilities of capture, we add a factor proportional to the number of singletons to the equiprobable estimate. 
We note that a high \emph{estimated} coefficient of variation combined with a high number of singletons might result in significant overestimation.

\begin{equation}
\label{eq:chao}
    \hat{N}_{\textsc{chao92}} = 
    \frac{D}{\hat{S}} +
    \frac{n(1-\hat{S})}{\hat{S}} 
    \gamma^2 =
    \frac{D + f_1 \gamma^2}{\hat{S}}
\end{equation}

\subsection{Evaluation Metrics}
\label{sec:metrics}

We evaluate the robustness and convergence of our estimators using the following metrics.

\subsubsection{Error Metric}
To evaluate the performance of the estimators in a controlled setting, we leverage the error metric introduced in~\cite{trushkowsky2013crowdsourced}.
For reference, the $\phi$ error metric aims at capturing the bias of the estimates as the absolute distance from the ground truth, if available. 
The sample order weighs the bias terms, that is, more recent errors get penalized more heavily. 
Conducting such an evaluation requires the ground truth value of the class size $N$, as well as the estimates calculated on the time-ordered sample periods.

\begin{equation}
    \phi = \frac{\sum_{i=1}^{k} \left|\hat{N_i} - N\right| i}{\sum i} = \frac{2 \sum_{i=1}^{k} |\hat{N_i} - N|}{k(k+1)}
\end{equation}

\subsubsection{Convergence Metric}
Conversely, we introduce a new metric $\rho$ that aims at evaluating the convergence of a given estimate. This metric acts as the main measurement tool in a real scenario where we do not have access to the ground truth, e.g. when performing large-scale analyses of completeness across classes.
The metric is derived from  $\phi$, as we look for stability and close distance between the estimate and the number $D$ of distinct values.
In contrast to the error metric, only the last $w$ observed samples out of the full set of samples are used in the convergence metric.
The closer the metric is to zero, the more confident we are that the class has converged to its complete set.
\begin{equation}
    \rho = \frac{\sum_{i=k-w}^{k} \frac{|\hat{N_i} - D_i|}{D_i}}{w}
\end{equation}

In the following section, we evaluate the presented estimators on a set of eight classes from Wikidata. We report our findings using the error and convergence metrics for the following estimators: Jack1 $(\hat{N}_{\textsc{jack1}}$)\footnote{We do not report on Jack2 as it has been shown to over-estimate the population size when the sample size is large~\cite{chiu2014improved}, which we have experienced as well.
}, N1-UNIF ($\hat{N}_{\textsc{N1-UNIF}}$), SOR ($\hat{N}_{\textsc{SOR}}$) and Chao92 ($\hat{N}_{\textsc{chao92}}$).
\section{Experimental Evaluation}
\label{sec:experiments}

We discuss the results of an extensive experimental evaluation of the estimators introduced in \cref{sec:model} below, starting with the description of the dataset we used. 
We obtain the full edit history of the Knowledge Graph and collect the observations for all the classes we found in Wikidata. We then selected a sub-sample of classes for which we have meaningful characteristics regarding the number of observations spread over time. From this set, we randomly selected classes and searched for an independent authoritative source that reports their true cardinality.
We set the sample period to 30 days, which results in at least one observation per sample period on most classes we selected.
We use the last four samples ($w = 4$ which equals roughly 4 Months) of our results to calculate the convergence metric.
Note that if an instance was not assigned the correct class we are not able to count it and we consider it as missing.
This is a desirable effect since a declarative query on Wikidata requesting the full list of a class will not return such instances either.

\subsection{Data}
\label{sec:data}

To evaluate our class completeness estimation methods, we use two different datasets from Wikidata: First, we use the \emph{entity graph}, provided by the Wikidata JSON dumps as of Aug 18, 2018\footnote{JSON Dump: \url{https://doi.org/10.5281/zenodo.3268725}}. The JSON dump contains the actual node descriptions and the edges between the nodes. 
Second, we use \emph{the edit history} as of Oct 1, 2018 provided in the Wikibase XML Dump\footnote{Edit History: \url{https://dumps.wikimedia.org/wikidatawiki/latest/}}. The edit history provides the list of all actions performed on the KG including the creation of new items, the update of labels and other values, as well as reverted edits\footnote{List of all Wikibase actions: \url{https://www.mediawiki.org/wiki/Wikibase/API/en}}. For each action, the XML dump provides the item changed, the user who made the change, the timestamp, a comment describing the action, and a pointer to the state of the graph before this action. 

\subsubsection{Dataset Description: Entity Degree Distribution}
\begin{figure}[th!]
\begin{center}
\includegraphics[width=1\textwidth, trim=1.9cm 0.3cm 2.5cm 1.3cm, clip=true]{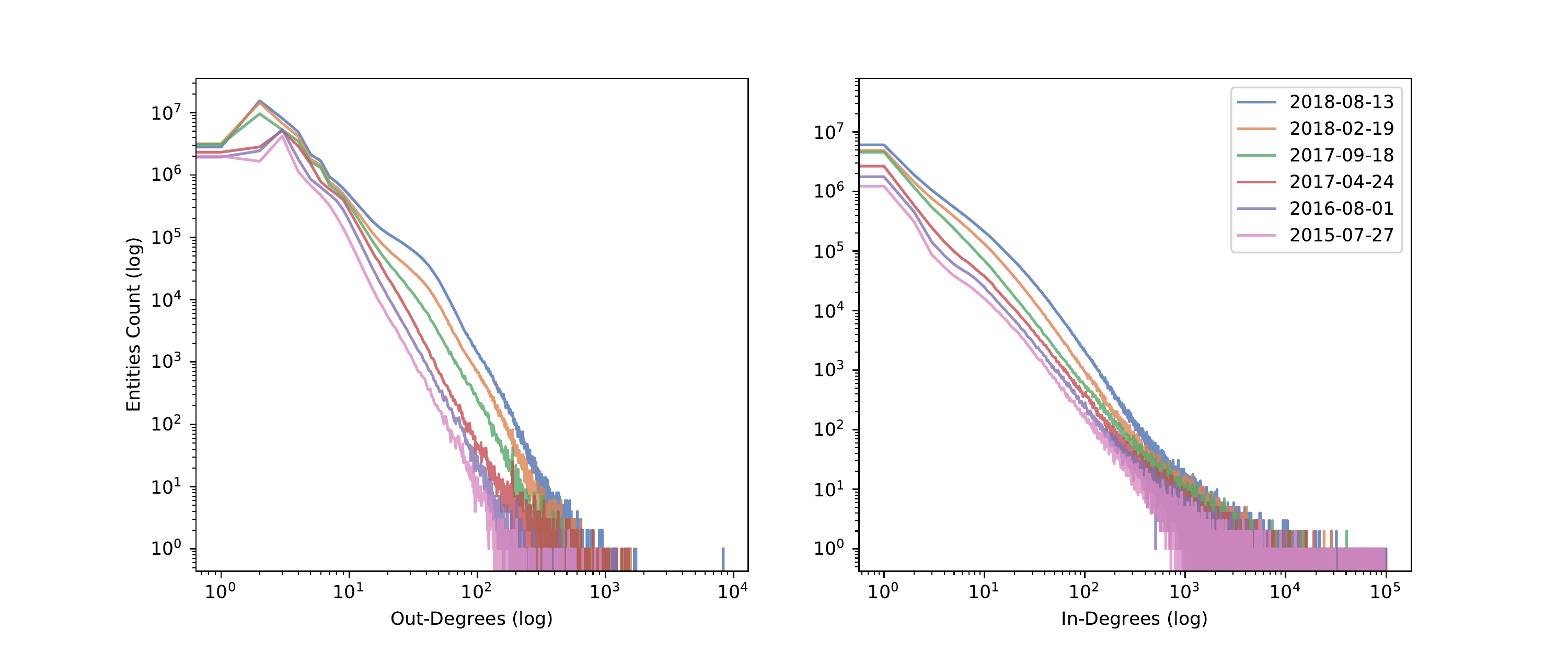}
\end{center}
\caption{The evolution of Wikidata: a temporal view on how the in- and out-degree distributions have evolved since the inception of the project.}
\label{fig:evol}
\end{figure}

To explore the characteristics of the dataset, we look at the graph as a whole (Fig. \ref{fig:evol}) and observe the constant overall growth of entities with different in and out-degrees at different points in time.

\subsubsection{Classes and Instances in the case of Wikidata}
\label{sec:classes}
Wikidata can be interpreted as an RDF graph \cite{DBLP:conf/semweb/ErxlebenGKMV14}, with a data model that differentiates between entities (including classes and instances) and properties. 
We define classes and instances in the Wikidata graph $G = (V, E)$ as follows:

\paragraph{Single Domain Classes}
In Wikidata, edges with the explicit label $E_{P31}$: instanceOf and $E_{P279}$: subclassOf explicitly define classes.\footnote{\url{https://www.wikidata.org/wiki/Wikidata:WikiProject_Ontology/Classes}} The target vertex $V_t$ which can be reached by following the edge with label $E_{P31}$: instanceOf from the source vertex $V_s$ are part of the classes $C$. Super classes collect all instances of a class $C$ which follow the edge $E_{P279}$: subclassOf once or multiple times.

\begin{lstlisting}[caption={Retrieves all instances of a specified single domain.},label={lst:single},language=SPARQL]
SELECT ?instance
WHERE { ?instance wdt:P31/wdt:P279* wd:Q515. }
\end{lstlisting}
To extract all instances of class $C_{Q515}: City$ we issue Query~\ref{lst:single} against the Wikidata endpoint.

\paragraph{Composite Classes}
We create composite classes by joining a class $C$ on one or multiple properties $E$ and their target instances $V$.
As an example, we can join class $C_{Q515}$: City with property $E_{P17}$: country and target $V_{Q142}$: France on the instances of $C_{Q515}$. The result is a composite class of all Cities in France $C_{Q515 \bowtie P17,Q142}$.

\begin{lstlisting}[caption={Retrieves all instances of a composite class.},label={lst:adhoc},language=SPARQL]
SELECT ?instance
WHERE { ?instance wdt:P31/wdt:P279* wd:Q515.
        ?instance wdt:P17 wd:Q142. }
\end{lstlisting}
As an example, Query~\ref{lst:adhoc} selects instances from the aforementioned composite class $C_{Q515 \bowtie P17,Q142}$.

\subsubsection{Data preparation}
The massive size of the edit history made it impossible to extract all observations from a database efficiently. 
Thus, in a first step we pre-process the edit history.
We select all edits involving at least two entities $V$ which therefore could be used to extract observations. 
The resulting intermediate data provides more than 161 million edits containing the source entity $V_s$, the property label of the connecting $E$, the target entity $V_t$, as well as the timestamp and the user.
In a second step, we pre-processed the JSON Dump into an in-memory graph to get fast access to all instances $V$ and properties $E$ (with property labels) of the Wikidata Graph.
This gives us information on which entity $V$ belongs to which class $C$.
Finally, to extract the observations pointing to an entity, we join the Wikidata edits with the in-memory Graph.

We filter out the observations belonging to a specific class $C$ by joining the observations pointing to an entity which in turn point to a class.
The resulting data, grouped by class, consists of 370 million distinct observations.

\subsection{Results}
\label{sec:results}

\begin{figure}[th!]
\centering
    \includegraphics[width=\textwidth, trim=3.3cm 3cm 3.5cm 2.8cm, clip=true]{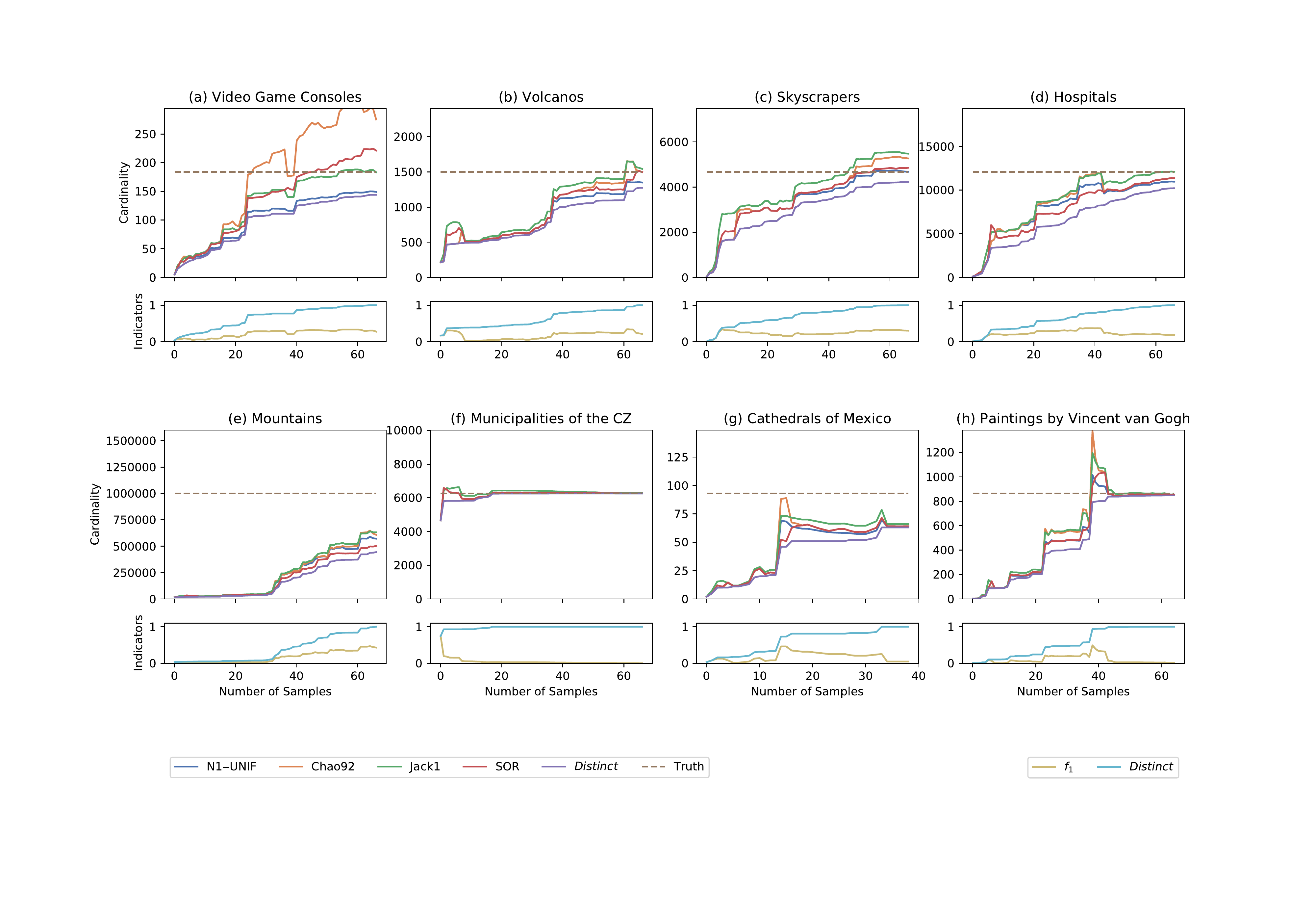}
\caption{Estimators used on Single Domain (a)-(f) and Composite classes (g)-(h).}
\label{fig:gt_graphs1}
\end{figure}

\Cref{fig:gt_graphs1} shows the results of the various estimators we consider.
The top part of each plot represents the results of the estimators for a specific domain, as well as the lower bound given by the absolute number of \emph{distinct} instances observed.
The x-axis represents the number of sample periods that we obtain in chronological order to perform the class size estimation.
At each sample period, we run an estimator using all the data samples collected so far.
The dashed line indicates the ground truth size.
The bottom part of each plot shows a comparison of two indicators: \emph{Distinct}, the distinct number of instances up to the sample period and \emph{$f_1$}, the proportion of instances observed exactly once, both normalized to the distinct number of instances retrieved in the end.
These indicators are key to our methods and serve the purpose of explaining the behavior of each estimator with respect to the properties of the samples.
In the following, we discuss these results and highlight key properties of each set of observations.

\subsubsection{Size Estimates on Single Domain Classes}

First, we inspect the results of estimating the size of a class when the query involves a single class definition.
\begin{table*}[!ht]
\centering
\caption{Performance evaluation of the estimators compared to the lower bound of the count of distinct instances. For each estimator we report the error $\phi$ and the convergence $\rho$. Results in bold indicates the lowest error for a given estimator. $N$ is the groundtruth and $D$ is the number of distinct instances on the last sample.}
\scalebox{1}{
\begin{tabularx}{\textwidth}{lrrrrrr}
\toprule
{}                                   &          &   \textbf{N1-UNIF} &    \textbf{Chao92} &      \textbf{Jack1} &   \textbf{SOR} & Distinct    \\
\midrule
\textbf{Video Game Consoles}         &  $\phi$  &      57.7 &      79.3 &\textbf{27.4} &       36.0 &       64.7    \\
($N = 184, D = 144$)                   &  $\rho$  &    0.0403 &    1.0096 &     0.2929 &     0.5529 & \vspace{0.14cm}  \\
\textbf{Volcanos}                    &  $\phi$  &     468.3 &     395.4 &\textbf{339.7} &      415.8 &      550.2    \\
($N = 1500, D = 1273$)                 &  $\rho$  &    0.0739 &    0.2300 &     0.2545 &     0.1700 & \vspace{0.14cm}  \\
\textbf{Skyscrapers}                 &  $\phi$  &     678.6 &     826.0 &      758.4 &      \textbf{650.4} &     1109.1    \\
($N = 4669, D = 4222$)                 &  $\rho$  &    0.1133 &    0.2560 &     0.3053 &     0.1482 & \vspace{0.14cm}  \\
\textbf{Hospitals}                   &  $\phi$  &     2,462 &     2,080 &\textbf{1,538} &    2,663 &    3945     \\
($N = 12090, D = 10215$)               &  $\rho$  &    0.0760 &    0.1126 &     0.1875 &     0.1126 & \vspace{0.14cm}  \\ 
\textbf{Mountains}                   &  $\phi$  &   671,874 &   656,653 &\textbf{643,616} &  709,178 &  751,938    \\
($N = 1000809, D = 444222$)            &  $\rho$  &    0.3255 &    0.4404 &     0.4503 &     0.1359 & \vspace{0.14cm}  \\
\textbf{Municipalities of the CZ}    &  $\phi$  &\textbf{22.2} &      31.3 &       86.3 &       31.3 &      26.6     \\
($N = 6258, D = 6256$)                 &  $\rho$  &    0.0002 &    0.0008 &     0.0029 &     0.0008 & \vspace{0.14cm}  \\   
\textbf{Cathedrals of Mexico}       &  $\phi$  &      37.2 &      35.0 &\textbf{31.7} &       36.6 &      43.1     \\
($N = 93, D = 63$)                     &  $\rho$  &    0.0159 &    0.0162 &     0.0463 &     0.0162 & \vspace{0.14cm}  \\
\textbf{Paintings by V. van Gogh}   &  $\phi$  &     184.8 &     183.1 &\textbf{173.0} &      189.1 &     204.9     \\
($N = 864, D = 848$)                   &  $\rho$  &    0.0027 &    0.0028 &     0.0119 &     0.0028 & \vspace{0.14cm}  \\
\bottomrule
\end{tabularx}
}

\label{tab:gt_tab}
\end{table*}
The first five figures show incomplete classes.
In Fig. \ref{fig:gt_graphs1}(a) we show the results for the small-sized class Video Game Consoles ($N=184$)\footnote{\scriptsize \url{https://en.wikipedia.org/wiki/List_of_home_video_game_consoles}}. We note how Chao92 is particularly intolerant to the small class size and overestimates.
\cref{fig:gt_graphs1}(b) shows the estimators for the class Volcanos ($N=1500$)\footnote{\scriptsize \url{https://www.usgs.gov/faqs/how-many-active-volcanoes-are-there-earth}}. Figure \cref{fig:gt_graphs1}(c), for Skyscrapers ($N=4669$)\footnote{\scriptsize \url{http://www.skyscrapercenter.com/}}, shows a class that is almost complete. The estimators are overshooting, because the $f_1$ on the available instances is high.
In \cref{fig:gt_graphs1}(d) Hospitals ($N=12090$)\footnote{\scriptsize \url{https://gateway.euro.who.int/en/indicators/hfa_471-5011-number-of-hospitals/}}, we observe how large classes also bring larger numbers of observations. This in turn helps the estimators to get stable before completeness is reached.
A massive class is represented with  \cref{fig:gt_graphs1}(e) Mountains ($N=1000809$)\footnote{\scriptsize \url{https://peakvisor.com/en/news/how_many_mountains_on_earth.html}}. We are aware that the ground truth, even if well researched by the source, is still rather suggestive. Nevertheless,  the estimators suggest that there are missing instances.
\textbf{}Finally,  \cref{fig:gt_graphs1}(f) Municipalities of the Czech Republic ($N=6258)$\footnote{\scriptsize \url{https://www.oecd.org/regional/regional-policy/Subnational-governments-in-OECD-Countries-Key-Data-2018.pdf}} shows a class which was complete early (around Sample 10). All estimators slowly converge to the ground truth.

\subsubsection{Size Estimates on Composite Classes}

As composite classes are by definition a subset of instances, compared to single domain classes, the associated observations can also drop to low numbers.
\Cref{fig:gt_graphs1}(g) shows such a case where the number of observations involving instances of a $C_{Q2977\bowtie P17,Q96}$ Cathedrals in Mexico ($N=93$)\footnote{\scriptsize \url{https://en.wikipedia.org/wiki/List_of_cathedrals_in_Mexico}} is $n=387$.
Figure \cref{fig:gt_graphs1}(h) $C_{Q3305213\bowtie P170,Q5582}$ Paintings by Vincent van Gogh ($T=864$)\footnote{\scriptsize \url{https://de.wikipedia.org/wiki/Vincent_van_Gogh#cite_note-Thomson_84-1}} is an example which displays the different phases of an estimator can encounter until class completeness. Starting by growing slowly at first with the addition of the first few elements. We observe intermittent overshooting when a large number of instances are added in a batch process. The final phase is a fast convergence towards the value of the ground truth.

\subsubsection{Performance Evaluation}
For all our experiments, we computed the error and convergence metrics introduced in \cref{sec:metrics} to obtain quantitative measurements on how the estimators perform and how they can be used. 
\cref{tab:gt_tab} summarizes the evaluation results across all classes considered in our work. We observe that Jack1 and SOR consistently achieve the lowest error rate across all classes.

\subsection{Discussion}
Our experimental results unveiled key properties in terms of the sensitivity and conditions under which some estimators perform better than others.
Generally speaking, all estimators beat the lower bound of distinct numbers in the error metric $\phi$. The exception is the class (Municipalities of the CZ) which converged early on, and for which N1-UNIF still beats the error of the distinct values. However, the other estimators lose against the lower bound (distinct) in this example on the number of instances because they over estimate the class size in the early samples before the class reaches completeness. We observe that more conservative estimators N1-UNIF, Chao92 perform worse then Jack1 and SOR for incomplete classes, which is why we recommend the last two in the end for the estimation of the class size.
The convergence metric can be used as an indicator to distinguish complete from incomplete classes without requiring the knowledge of the real class size.
In \cref{tab:gt_tab}, we see how the convergence metrics $\rho$ are low ($<0.001$) for complete classes. On the other hand for incomplete classes $\rho$ is comparatively high ($>0.1$).
\cref{tab:convergences} lists ten randomly-picked classes, along with the convergence on \emph{SOR} and the number of distinct instances, for a low and high $\rho$ values suggesting complete and incomplete classes respectively.
These lists illustrate how our convergence metric can be leveraged to identify gaps in the KG.

\begin{table}[t]
\caption{Lists of 10 randomly picked examples. Left with a low $\rho$ suggesting a complete class, and right a high $\rho$ suggesting an incomplete class.}
\scalebox{1}{

\begin{tabular}{lrr}
\toprule
\multicolumn{2}{r}{SOR $\rho < 0.001$} & Distinct \\
\midrule
municipality of Japan     &     0.0000 &      739 \\
Philippine TV series      &     0.0009 &      822 \\
Landgemeinde of Austria   &     0.0000 &    1,116 \\
district of China         &     0.0009 &      975 \\
nuclear isomer            &     0.0002 &    1,322 \\
international border      &     0.0000 &      529 \\
commune of France         &     0.0001 &   34,937 \\
village of Burkina Faso   &     0.0005 &    2,723 \\
supernova                 &     0.0005 &    5,906 \\
township of Indiana       &     0.0002 &      999 \\
\bottomrule
\end{tabular}

\hfill
\begin{tabular}{lrr}
\toprule
\multicolumn{2}{r}{SOR $\rho > 0.1$} & Distinct \\
\midrule
urban beach                    &     0.1759 &      683 \\
hydroelectric power station    &     0.2975 &    2,936 \\
aircraft model                 &     0.1800 &    3,919 \\
motorcycle manufacturer        &     0.1758 &      690 \\
local museum                   &     0.1760 &    1,150 \\
waterfall                      &     0.1942 &    5,322 \\
race track                     &     0.2783 &      946 \\
film production company        &     0.2107 &    2,179 \\
red telephone box              &     0.3469 &    2,716 \\
mountain range                 &     0.2390 &   21,390 \\
\bottomrule
\end{tabular}\textbf{}
}
\label{tab:convergences}
\end{table}

\subsection{Additional Material and Tools}
\label{sec:tools}

The results on all classes in Wikidata are available at \url{http://cardinal.exascale.info}.
We also release our Python code implementing the data processing pipeline, all estimators and metrics as an open source package\footnote{\url{https://github.com/eXascaleInfolab/cardinal/}}.
This includes tools to seek for incomplete classes based on the convergence metric.
Finally, we provide the pre-processed data at every step of the processing pipeline, as well as the final results for each dataset.
\section{Conclusions and Future Work}
\label{sec:conclusions}

In this work, we introduced a methodology to estimate class sizes in a collaborative KG and evaluated it over Wikidata. We showed how collaborative editing dynamics create a trove of information that can be mined to extract information about data access, edits, data linkage, and overall graph growth.
We relied on the edit history over six years of activity in Wikidata to collect capture-recapture observations related to a particular entity within the class of interest.
We reviewed, applied, and evaluated a battery of non-parametric statistical techniques that leverage frequency statistics collected from the data to estimate the completeness of a given class.

Our experimental results show that many estimators yield accurate estimates when provided with enough observations that reflect the actual underlying distribution of the instances of a class.
However, some estimators like Chao92 tend to be less robust to bursts of newly discovered instances.
Finally, based on our results, we provided a set of practical recommendations to use convergence metric in conjunction with  estimators to decide whether a particular class is complete or to perform large-scale completeness analyses.
%
Our work has direct implications for both Wikidata editors and data consumers. We can provide convergence statistics on the estimated class completeness by domains to point to knowledge gaps.
Such statistics could aid newcomers, who often feel insecure about what to edit, to decide what to contribute or what to focus on. 

In future work, we plan to leverage statistics on page views showing the attention that specific groups of items receive within the KG to inform estimators. 
We would also like to develop parametric models that assume a particular edit probability distribution.
This is especially applicable to domains with a considerable bias towards popular entities such as Humans and Musicians. 
Another area of potential development is the usage of completeness estimators to detect systematic errors in Wikidata: While exploring the data, we have  observed many cases of misclassification, which we conjecture as being the result of the growing complexity of the Wikidata ontology that includes more than forty thousand classes at the time of writing.

\section{Acknowledgements}
This project has received funding from the European Research Council (ERC) under the European Union's Horizon 2020 research and innovation programme (grant agreement 683253/GraphInt). It is also supported by the Australian Research Council (ARC) Discovery Project (Grant No. DP190102141).

\bibliographystyle{splncs04}
\bibliography{cardinal_iswc2019}

\end{document}